\documentclass[aps,prd,nofootinbib,showkeys,floatfix,twocolumn]{revtex4-1}
\pdfoutput=1
\usepackage{amsmath}
\usepackage{amssymb}
\usepackage{graphicx}
\usepackage[mathscr]{eucal}
\usepackage{color}
\usepackage[T1]{fontenc} 
\usepackage{slashed}
\usepackage{xspace}
\usepackage{hyperref}
\usepackage{cleveref}
\usepackage{xspace}
\usepackage{ulem}
\usepackage{cancel}
\usepackage{enumerate}
\usepackage{units}
\usepackage[utf8]{inputenc}

\newcommand\SARAH{{\tt SARAH}\xspace}
\newcommand\SPheno{{\tt SPheno}\xspace}
\newcommand\Fortran{{\tt Fortran}\xspace}

\newcommand{\DRbar}{{\ensuremath{\overline{\mathrm{DR}}}}\xspace}

\newcommand{\AddrCERN}{%
Theoretical Physics Department, CERN, Geneva, Switzerland
}

\begin{document}

\hfill  CERN-TH-2016-061 \vspace{0.2cm}

 \title{A closer look at non-decoupling $D$-Terms}

 \author{Florian Staub}
 \email{florian.staub@cern.ch}
 \affiliation{\AddrCERN}


\begin{abstract}
Non-Decoupling D-Terms are an attractive possibility to enhance the tree-level mass of the standard model like Higgs boson in supersymmetric models. We discuss here for the case of a new Abelian gauge group two effects usually neglected in literature: (i) the size of the additional radiative corrections to the Higgs mass due to the presence of the new gauge coupling, and (ii) the impact of gauge kinetic mixing. It is shown that both effects reduce to some extent the positive effect of the non-decoupling D-terms on the Higgs mass.

\end{abstract}
\maketitle

\section{Introduction}
\label{sec:intro} 
The discovery of the Higgs boson at the Large Hadron Collider (LHC) \cite{Chatrchyan:2012ufa,Aad:2012tfa} 
completes not only the Standard Model (SM) of particle physics, but gives also new constraints on any extension of it. 
The most studied extension of the SM to date is supersymmetry (SUSY), see Ref.~\cite{Nilles:1983ge,Martin:1997ns} and references therein. 
However, the minimal-supersymmetric standard model (MSSM) is facing increasing pressure over the last few years. One of the reasons 
is the measured Higgs mass of 125~GeV which is not trivial to accommodate in the MSSM. The tree-level mass of the Higgs has the strict upper
limit $m_h^T \le M_Z$, and one  therefore needs large radiative corrections to push the mass to the required value. The two main possibilities 
within the MSSM are either heavy stops, and/or a large mixing among them. The first option raises immediately the question about 
naturalness \cite{Batell:2015fma}, while the second one can cause charge and colour breaking minima \cite{Camargo-Molina:2013sta,Camargo-Molina:2014pwa,Chattopadhyay:2014gfa}. \\
For these reasons the interest in non-minimal SUSY models has increased in the last years. A widely studied approach to soften the need for large radiative corrections 
to the Higgs mass is to increase the mass already at tree-level. This is done for instance in different singlet extensions via $F$-terms 
\cite{BasteroGil:2000bw,Dermisek:2005gg,Ellwanger:2009dp,Ellwanger:2006rm,Delgado:2010uj,Ross:2011xv,Hall:2011aa}. The other option is to use $D$-terms \cite{Ma:2011ea,Zhang:2008jm,Hirsch:2011hg}. \\
The simplest realisations of these ideas show the expected decoupling behaviour: as soon as the new sector responsible for the enhanced tree-level 
mass becomes heavy, the positive effect of the Higgs mass shrinks. However, it has been shown that this decoupling can be circumvented at the price of 
introducing large soft-breaking masses. Indeed, valid models with non-decoupling $F$- \cite{Lu:2013cta,Kaminska:2014wia,Ding:2015wma} 
or $D$-terms \cite{Batra:2003nj,Maloney:2004rc,Babu:2004xg,Martinez:2004rh,Bertuzzo:2014sma,Capdevilla:2015qwa} have been successfully constructed. \\
We will consider in this letter the case of non-decoupling $D$-terms via a new Abelian gauge group $U(1)_X$. So far, only the leading order effect on the Higgs 
mass was studied in this scenario: the impact of the non-decoupling $D$-terms at tree-level in the case of vanishing gauge kinetic mixing was considered. We will show that (i) radiative 
corrections and (ii) the presence of gauge-kinetic mixing can give important corrections to this approximation. \\
This letter is organised as follows: in \cref{sec:gkm} a short excursion into gauge kinetic mixing in supersymmetric models is given, before in \cref{sec:model} the origin of both effects 
in a concrete model is described. The impact on the Higgs mass is analysed numerically in \cref{sec:numerics}, before we summarise in \cref{sec:summary}.

\section{Gauge kinetic mixing in supersymmetric models}
\label{sec:gkm}
If more than one Abelian gauge group is present, mixing terms between 
the field strength tensors of the groups are allowed by gauge and Lorentz invariance \cite{Holdom:1985ag}:
\begin{equation}
\label{eq:off}
\mathcal{L} = -\frac{1}{4} \kappa F_{\mu\nu}^A F^{B,\mu\nu}, \hspace{1cm} A\neq B \, ,
\end{equation}
Here, $F^{\mu\nu}$ are the field strength tensors of two different
Abelian groups $A$, $B$ and $\kappa$ is in general
an $n \times n$ matrix if $n$ Abelian groups are present. For practical 
applications it is often useful to bring the propagators of the 
vector bosons back into their canonical form by the redefinition \cite{Fonseca:2013bua}
\begin{equation}
V\to \kappa^{1/2}V\,.
\end{equation}
The trade-off for this redefinition is  a change in the 
covariant derivative:
\begin{equation}
\partial_\mu - i Q_i^{T}{\tilde G} V \to \partial_\mu - i  Q_i^{T}{\tilde G}\kappa^{-1/2}V\,,
\end{equation}
where $\tilde G$ is the original diagonal matrix of $n$ individual gauge couplings
associated to the $n$  Abelian gauge factors,  and  $Q_i$ is the
vector of the
relevant $U(1)$ charges. 
Thus, the $\kappa^{-1/2}$ factor can  be translated into a new set of  $\tfrac{1}{2}n(n-1)$ ``effective'' 
gauge couplings whose combinations populate off-diagonal entries of a  gauge-coupling matrix  
\begin{equation}
\label{Gmatrix}
G\equiv \tilde G \kappa^{-1/2}\,,
\end{equation}
In supersymmetric models, gauge kinetic mixing appears not only for the  vector bosons, but also for 
both, the $D$-terms and the gaugino soft-breaking terms. Of particular importance in the following are the 
new $D$-terms which read for the  Abelian sector
\begin{equation}
\label{eq:Dabelian}
\mathcal{L}_{D,U(1)} = \sum_{ij} (\phi_i^*\phi_i) (G^T Q_{i}) (G Q_j) (\phi_j^* \phi_j)
\end{equation}
while the non-Abelian $D$-terms remain unchanged. \\
For the special case of two Abelian gauge groups, the most general form of $G$ reads
\begin{equation}
 G=\left(\begin{array}{cc} 
          g_{XX} & g_{XY} \\ g_{YX} & g_{YY}
         \end{array}\right)
\end{equation}
In general, one has the freedom to perform a change in basis to bring $G$ into a particular form. 
The most commonly considered cases are the symmetric basis with $g_{XY} = g_{YX} = \tilde{g}$ and the triangle basis with $\tilde{g} = g_{XY}$, $g_{YX}=0$.  
The triangle basis has the advantage that the new scalars do not contribute to the electroweak VEV, and the entire impact of gauge kinetic mixing is encoded in one new coupling $\tilde{g}$. The relations between $g_{ij}$ ($i,j=X,Y$) and the physical couplings called $g_Y$, $g_X$, $\tilde{g}$ are given by \cite{O'Leary:2011yq} 
\begin{eqnarray}
& g_Y = \frac{g_{YY} g_{XX} - g_{XY} g_{YX}}{\sqrt{g_{XX}^2 + g_{XY}^2}}\,, \quad\quad
g_X = \sqrt{g_{XX}^2 + g_{XY}^2}\,, &  \nonumber \\
\label{eq:basis}
& \tilde{g} = \frac{g_{YX} g_{XX} + g_{YY}  g_{XY}}{\sqrt{g_{XX}^2 + g_{XY}^2}} \,. &
\end{eqnarray}

\section{A model for non-decoupling $D$-terms}
\label{sec:model}
\subsection{The Higgs mass at tree-level}
In order to show the impact of loop corrections and gauge kinetic mixing on the Higgs mass in the presence of non-decoupling $D$-terms, we take a model inspired by Ref.~\cite{Capdevilla:2015qwa} as example\footnote{In contrast to Ref.~\cite{Capdevilla:2015qwa} we changed the $U(1)_X$ charges of the new fields to allow for a seesaw type-I via the term $Y_x\,\hat{\nu}\,\hat{\bar{\eta}}\,\hat{\nu}$}. 
In this model, the particle content of the MSSM is extended by a gauge singlet superfield $\hat S$, a superfield for the right-handed neutrino and two fields $\hat \eta$,$\hat\bar\eta$ responsible for the breaking of $U(1)_X$. It has also been shown that this model, when extended in addition by three generations of vector-like leptons and quarks, provides an attractive explanation for the diphoton excess observed at 750~GeV by ATLAS and CMS \cite{Staub:2016dxq}.
\begin{table}[h]
\centering
\begin{tabular}{|c|cccccc|ccccc|} 
\hline 
   & $\hat q$ & $\hat l$ & $\hat d$ & $\hat u$ & $\hat e$ & $\hat \nu$ & $\hat H_d$ &$\hat H_u$  & $\hat{\eta}$ & $\hat{\bar{\eta}}$ & $\hat{S}$ \\
\hline 
$U(1)_X$ & 0 & 0 &  $\frac{1}{2} $ & $-\frac{1}{2} $  & $\frac{1}{2} $  & $-\frac{1}{2} $ & $-\frac{1}{2} $  & $\frac{1}{2} $  & -1 & 1 & 0 \\
\hline
\end{tabular} 
\caption{Charges of the superfields under $U(1)_X$. The superfields in addition to the MSSM particle content are singlets under the SM gauge group.}
\label{tab:U1xMSSMparticles}
\end{table}
The $U(1)_X$ charges of all superfields are summarised in \cref{tab:U1xMSSMparticles}, while the most general superpotential in agreement with all global and local symmetries is given by
\begin{align} 
\nonumber W &= W_{MSSM} +Y_x\,\hat{\nu}\,\hat{\bar{\eta}}\,\hat{\nu} +(\mu + {\lambda} \hat{S})\,\hat{H}_u\,\hat{H}_d \nonumber \\ 
&\phantom{={}} + \hat{S} (\xi\, +  {\lambda}_X\,\hat{\eta}\,\hat{\bar{\eta}}) + M_S\,\hat{S}\,\hat{S} +\frac{1}{3} \kappa \,\hat{S}\,\hat{S}\,\hat{S} \,.
\end{align} 
For simplicity, we will neglect in the following $\lambda$, $M_S$, and $\kappa$ as well as their soft-breaking terms. In addition, the soft-breaking term of $\xi$ can always be
shifted away. The relevant soft-breaking terms in the scalar sector are
\begin{equation}
- \mathcal{L}_{SB} = \dots  + (B_\mu H_u H_d + T_\lambda S \eta \bar\eta + \text{h.c.}) + m_{\varphi}^2 |\varphi|^2  \,,
\end{equation}
with $\varphi= \{H_d,H_u,\eta,\bar\eta,S\}$. \\
Electroweak symmetry breaking and the breaking of $U(1)_X$ are triggered by the following scalars receiving vacuum expectation values (VEVs)
\begin{eqnarray} 
& H_d^0 =  \, \frac{1}{\sqrt{2}} \left( \phi_{d}  + v_d  + i  \sigma_{d} \right)\,, \quad 
H_u^0 =  \, \frac{1}{\sqrt{2}} \left(\phi_{u}  + v_u  + i  \sigma_{u} \right) \,, & \nonumber \\ 
& \eta =  \, \frac{1}{\sqrt{2}} \left(\phi_{\eta}  +  v_{\eta}  + i  \sigma_{\eta}\right)\,, \quad
\bar{\eta} =  \, \frac{1}{\sqrt{2}} \left( \phi_{\bar{\eta}}  + v_{\bar{\eta}}  + i  \sigma_{\bar{\eta}}\right) \,. & \nonumber \\ 
\end{eqnarray}
In addition, a VEV for the scalar singlet is induced in general
\begin{equation}
 S =  \, \frac{1}{\sqrt{2}} \left( {\phi}_{s}  + v_S  + i  {\sigma}_{s} \right) \, . 
\end{equation} 
We define $\tan\beta=\frac{v_u}{v_d}$, $v=\sqrt{v_d^2+v_u^2}$ as well as $\tan\beta_x = \frac{v_{\eta}}{v_{\bar \eta}}$, and $x=\sqrt{v_\eta^2+v_{\bar \eta}^2}$. One can use the five tadpole equations
\begin{equation}
\frac{\partial V}{\partial v_i} = 0 \hspace{1cm} i=u,d,S,\eta,\bar\eta 
\end{equation}
to eliminate five free parameters. We are doing this in the following for $m_{H_d}^2$, $m_{H_u}^2$, $m_S^2$, $m_{\eta}^2$ and $\xi$. Thus, the remaining free parameters in the scalar sector are $B_\mu$, $\mu$, $m_{\bar \eta}$, $\lambda$, $T_\lambda$ and the VEVs. In all numerical studies we are going to fix $\mu = \sqrt{B_\mu} = T_\lambda/\lambda= 1$~TeV, and $\lambda=-0.2$. 
The neutral Higgs mass matrix can be written in the basis $(h,H,N,N',S)$ as
\begin{equation}
\left(
\begin{array}{ccccc}
 \frac{g^2 v^2}{4} & \frac{g^2 v^2}{2 t_\beta} & 0 & \frac{1}{2} g_X g_T v x & 0 \\
 \bullet & t_\beta B_\mu+\frac{g^2 v^2}{t_\beta^2} & 0 & \frac{g_X g_T v x}{t_\beta} & 0 \\
 \bullet & \bullet & \frac{\lambda^2 x^2}{2} & 0 & \lambda ^2 x v_S-\frac{x T_\lambda}{\sqrt{2}} \\
 \bullet & \bullet & \bullet & m_{44} & 0 \\
 \bullet & \bullet & \bullet & \bullet & \frac{x^2 T_\lambda}{2 \sqrt{2} v_S} \\
\end{array}
\right) 
\end{equation}
with the abbreviations
\begin{equation}
m_{44} =  2 m^2_{\bar\eta} + \lambda^2 v_S^2+  \frac{1}{2} g_X \left(2 g_X x^2-\frac{g_X v^2}{t_\beta^2}+ g_T v^2 \right)
\end{equation}
and $g_T = g_X + \tilde{g}$, $g^2= g_1^2 + g_2^2 + g_T^2$. We have used the relations $h =\cos\beta \phi_d + \sin\beta \phi_u$,  $H =\cos\beta \phi_u - \sin\beta \phi_d$,$N =\cos\beta_x \eta + \sin\beta_X \bar\eta$,  $N' =\cos\beta_x \bar\eta- \sin\beta_X \eta$, and took the limit $\tan\beta \gg 1$, $\tan\beta_x =1$.  The lightest eigenstate of the $(h,N')$ sub-system is approximated by
\begin{equation}
m_h^2 \simeq \frac{\Big(2 \Big(g_{1}^{2} + g_{2}^{2} + \Big(\tilde g + g_{X}\Big)^{2}\Big)m_{\bar{\eta}}^2  + \Big(g_{1}^{2} + g_{2}^{2}\Big)M_{Z'}^{2} \Big)v^{2}}{4 \Big(2 m_{\bar{\eta}}^2  + M_{Z'}^{2}\Big)} \,.
\end{equation}
Here, we made use of $x = \frac{\sqrt{4 M_{Z'}^2 - (\tilde{g} + g_X)^2 v^2}}{g_x}$ and assumed $M_{Z'} \gg v$. In addition, we replaced $g_T$ as well as $g^2$ for clarity by the full expressions. 
We find the expected limits 
\begin{itemize}
 \item $M_{Z'} \gg m_{\bar{\eta}}$: $m_h = \frac{1}{4} (g_1^2 + g_2^2) v^2$
 \item $M_{Z'} \ll m_{\bar{\eta}}$: $m_h = \frac{1}{4} \left(g_1^2 + g_2^2 + (\tilde g + g_{X})^{2} \right) v^2$
\end{itemize}
Thus, we have the desired $D$-term enhancement for very large soft-terms, which is, however, modified by the presence of the off-diagonal gauge coupling $\tilde g$. 

\subsection{The running of gauge kinetic mixing}
\begin{figure}[tb]
\includegraphics[width=0.75\linewidth]{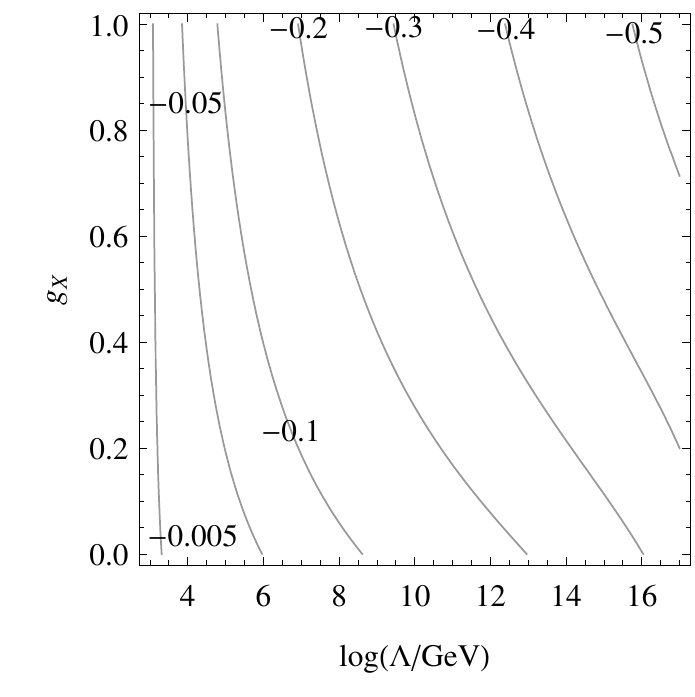}
\caption{$\frac{\tilde{g}}{g_X}$ at 1~TeV as function of the scale $\Lambda$ where gauge kinetic mixing vanishes and $g_X$ at the scale.}
\label{fig:gkm}
\end{figure}

As long as $\tilde g$ is small, the impact of gauge kinetic mixing will be negligible. Thus, as long as $\tilde g$ is treated as free parameter, it can be set to zero at a given scale. However, via renormalisation group effects (RGE) it will be induced radiatively if the two Abelian gauge groups are not orthogonal. Orthogonality in this respect means that the anomalous dimension matrix defined by
\begin{equation}
16 \pi^2 \gamma_{ab} = \text{Tr} Q_a Q_b
\end{equation}
is diagonal. Here,  $a$ and $b$ run over all $U(1)$ groups and the trace runs over all superfields charged under the corresponding group. In many commonly studied $U(1)$ extension as $\mathcal{G}_{SM} \times U(1)_{B-L}$, $\gamma$ is not diagonal \cite{Basso:2010jm}. Even if the two Abelian groups can be embedded in a singlet GUT group, gauge kinetic mixing might appear for the remaining {\it light} degrees of freedom \cite{Hirsch:2012kv,Krauss:2013jva,Athron:2015vxg}.
Calculating the anomalous dimension matrix for the model under consideration, one finds large off-diagonal entries 
\begin{equation}
16 \pi^2 \gamma = N \left(\begin{array}{cc} 11 & 7 \\ 7 & 9 \end{array} \right) N \,,
\end{equation}
where $N$ is a diagonal matrix containing the GUT normalisation. Since we don't assume any GUT embedding here, we use $N=\text{diag}(1,1)$ in the following. At the one-loop level one can calculate the value of $\tilde{g}$ at the SUSY scale as function of (i) the scale $\Lambda$ where $\tilde g$ is assumed to vanish and (ii) the value for $g_X$ at this scale. The running of the gauge coupling matrix $G$ at one-loop is given by \cite{delAguila:1988jz,Fonseca:2011vn}
\begin{equation}
\beta^{(1L)}_G \propto G G^T \gamma G
\end{equation}
One has to ensure to bring $G$ to the particular basis defined in eq.~(\ref{eq:basis}) after the running. The result is shown in \cref{fig:gkm}.
One finds that $\tilde{g}$ is in general negative, and only for a very low cut-off $\Lambda$ is the new coupling $\tilde g$ is much smaller than $g_X$. For $\Lambda = 10^6$~GeV and large $g_X$, $|\tilde g|$ can already be as large as 0.1, while for a GUT motivated scenario with $\Lambda \simeq 10^{16}$ and $g_X(\Lambda) = 0.72$, $\tilde g$ can even be as large as 0.4--0.5$\times g_X$ \footnote{Even if small values of $\tilde g$ are considered, it can be wrong to neglect gauge-kinetic mixing. Although the impact on the Higgs mass might not be  important, crucial effects for instance concerning $Z'$ searches \cite{Basso:2010pe,Krauss:2012ku} or dark matter properties \cite{Basso:2012gz} could be missed.}. For completeness, we want to mention that the possibility of positive ratios $\tilde g/g_X$ exists: this can appear for very large, but negative values of $g_X$ at the scale $\Lambda$.

\subsection{Loop corrections}
\begin{figure}[tb]
\includegraphics[width=1.1\linewidth]{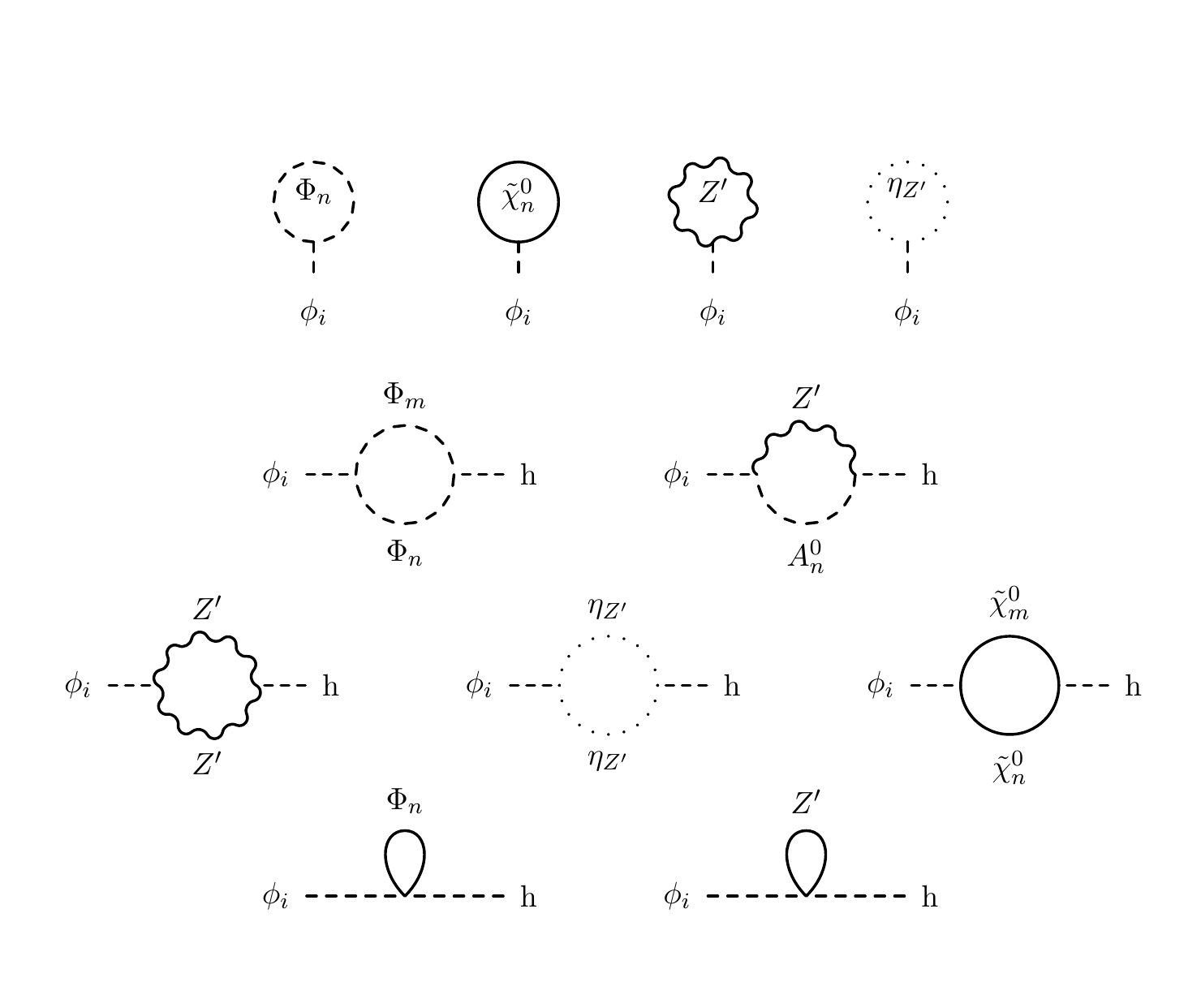}
\caption{One-loop diagrams contributing to the radiative corrections to $m_h$ proportional to $g_X^2$. The scalar contributions are summarised by $\Phi=\{h,A^0,H^+\}$, but for the case 
where only pseudo-scalars ($A^0$) can appear. $m,n$ label the generations in the loop.}
\label{fig:diagrams}
\end{figure}

\begin{figure}[tb]
\includegraphics[width=1.0\linewidth]{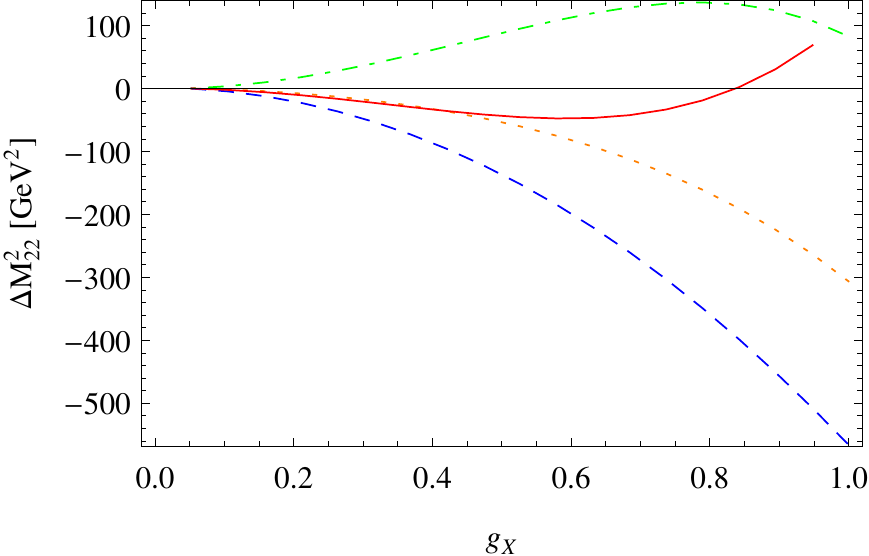}
\caption{Size of $\Delta M^2_{22}$ as function of $g_X$ for the diagrams involving only (i) neutralinos (green, dot-dashed), (ii) CP-even scalars (blue, dashed), (iii) charged Higgs (orange, dotted), (iv) CP-odd scalars, $Z'$ and its ghost (red, full).}
\label{fig:Pis}
\end{figure}

Because of the non-vanishing charge of the Higgs doublets under the new gauge group, many additional radiative contributions to the Higgs mass arise already at one-loop. The Feynman diagrams with possible contributions proportional to $g_X$ are depicted in \cref{fig:diagrams}. Since the right-sfermions carry also a $U(1)_x$ charge, they give also non-vanishing diagrams at one-loop. However, it can be shown that the tadpole and self-energy corrections cancel exactly for them. 
The loop corrected scalar masses are the eigenvalues of the radiatively corrected mass matrix $m_H^{2,(1L)}$ given by
\begin{equation}
m_H^{2,(1L)}(p^2) = m_H^{2,(T)} + \Delta M^2(p^2) 
\end{equation}
where $\Delta M$ is the sum of tadpole- and self-energy contributions. The pole masses $m_{h_i}^2$ are then associated with the eigenvalues of $m_H^{2,(1L)}(p^2=m_{h_i}^2)$. \\
For the calculation of the diagrams it is necessary to diagonalise the mass matrices of the particles in the loop, which have dimensions up to $8\times 8$. Thus, most contributions can only be calculated numerically. Only for the charged Higgs contribution we can find an analytical estimate for the impact on the SM-like Higgs mass which is mainly given by the change in the (2,2)-element of $\Delta M^2$:
\begin{align}
& \delta \Delta M^2(p^2) = \Delta M^2(p^2)|_{g_X=0} - \Delta M^2(p^2)|_{g_X \neq 0} \nonumber \\
=& \frac{g_X^2 v^2 \left(\left(g_X^2-2 g_2^2\right) \log \left(\frac{t_\beta B_\mu}{Q^2}\right)+\left(2 g_2^2+g_X^2\right) \log\left(\frac{M_W^2}{Q^2}\right)\right)}{64 \pi ^2 }
\end{align}
$Q^2$ is the renormalisation scale and we used the limit $p^2\to0$.  
In order to give an impression of the importance of all other loop corrections, we group all diagrams into four sets: (i) diagrams involving only neutralinos ($\tilde \chi^0$), (ii) diagrams involving only neutral CP-even scalars ($h$), (iii) diagrams involving only charged Higgs scalars ($H^+$), (iv) diagrams with CP-odd scalars, the new gauge boson and the corresponding ghost ($A^0$, $Z'$, $\eta'$). One has to consider the fields in the last group together, in order to have a gauge invariant set of diagrams. In \cref{fig:Pis} we show the values of the corrections to the (2,2)-element of the Higgs mass matrix as function of $g_X$ for the case of vanishing kinetic mixing, $M_{Z'} = 3$~TeV and $m_{\bar \eta} = 2$~TeV. 
One can see that the main loop corrections come from the charged and CP-even scalar sector. Since both are negative, one can expect that the new loop corrections reduce the SM-like Higgs mass, i.e. they might compensate to some extent the positive effect at tree-level. The corrections stemming from neutralinos are comparable small, while the ones involving the $Z'$, its Goldstone boson and its ghost are even smaller. 

\begin{figure}[tb]
\includegraphics[width=1.0\linewidth]{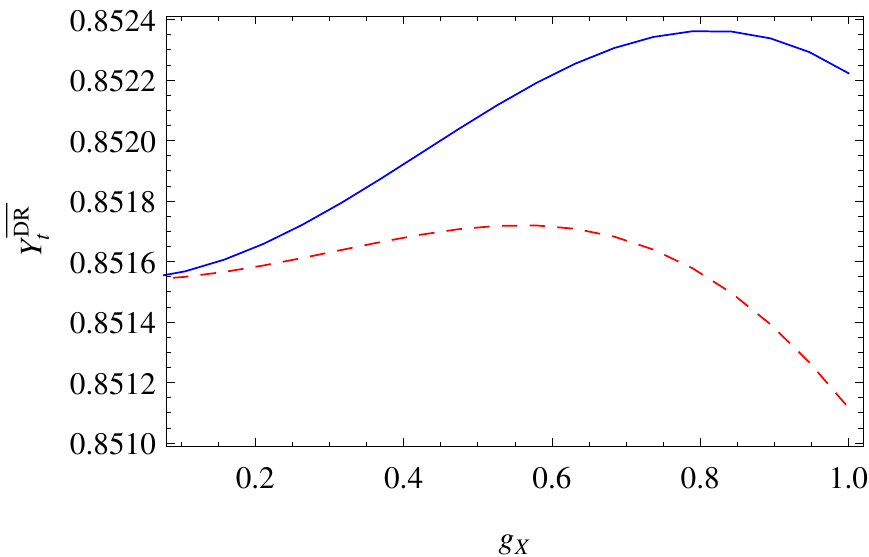}
\caption{Running top Yukawa coupling $Y_t^\DRbar$ at the SUSY scale $Q=1.5$~TeV as function of $g_X$. The red, dashed line is for $M_{Z'}=3$~TeV and the blue, full line for $M_{Z'}=4$~TeV. }
\label{fig:Yt}
\end{figure}

There is also an indirect effect of the extended gauge sector on the MSSM-like corrections to the Higgs mass: the new force changes the relation between the measured SM parameters and the running \DRbar values which enter the loop calculations. The most important parameter in this respect is the top Yukawa coupling $Y_t$. As it can be seen from \cref{fig:Yt}, the change in the running coupling $Y_t^\DRbar$ is only very small when turning up $g_X$. Thus, the change in the loop corrections is only very tiny.

\section{Numerical results}
\label{sec:numerics}
\begin{figure}[tb]
\includegraphics[width=1.0\linewidth]{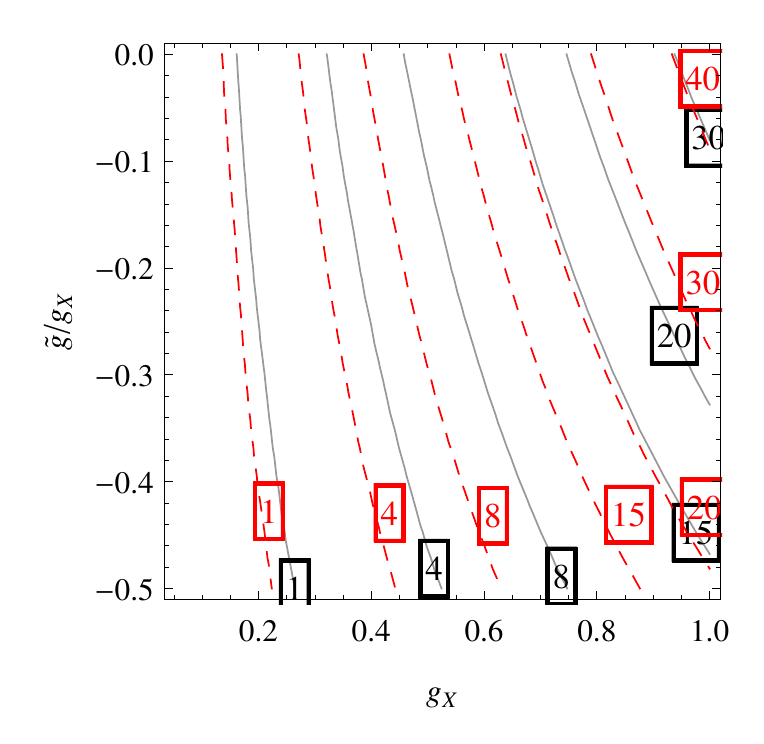} 
\caption{The enhancement of the SM-like Higgs mass (contours of constant shift in GeV) via the new gauge contributions at the tree-level  (red, dashed lines) and when including loop-effects (black, full lines). }
\label{fig:gXgTmh}
\end{figure}

\begin{figure}[tb]
\includegraphics[width=1.0\linewidth]{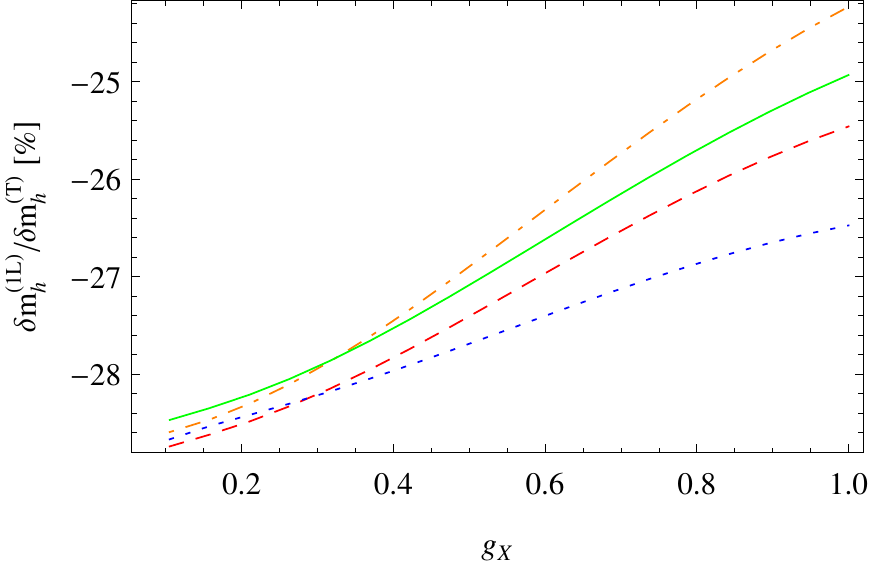} 
\caption{The relative size $\delta m_h^{(1L)}/\delta m_h^{(T)}$ as defined in \cref{eq:d1,eq:d2} as function of $g_X$. Here, we put $\tilde g =0$ and assumed 
$(M_{Z'},m_{\bar\eta})=(3,3)$~TeV [red, dashed],$(3,4)$~TeV [orange, dot-dotted],$(4,3)$~TeV [blue, dotted],$(4,4)$~TeV [green,full].}
\label{fig:Rel_T_L}
\end{figure}

We turn now to the discussion of the impact on the SM-like Higgs mass of the two effects described in the last section. For this purpose, we make use of the \SPheno \cite{Porod:2011nf,Porod:2003um} output of \SARAH \cite{Staub:2008uz,Staub:2009bi,Staub:2010jh,Staub:2012pb,Staub:2013tta,Staub:2015kfa}\footnote{The model files called {\tt U1xMSSM} are now included in the public \SARAH version}. The \SARAH generated \Fortran code provides a fully fledged spectrum generator for the model which enables a calculation of the entire mass spectrum at the one-loop level without any approximation: any loop contribution and the full momentum dependence is included. In principle, \SPheno includes also new two-loop corrections for the model under consideration  \cite{Goodsell:2014bna,Goodsell:2015ira}. However, these are so far only available in the gaugeless limit, i.e. the corrections we are mainly interested in are not covered. \\
We show in \cref{fig:gXgTmh} the shifts of the SM-like Higgs mass at tree-level and at the  one-loop level for $M_Z'=m_{\bar \eta}=3$~TeV in the $(g_X, \tilde g/g_X)$-plane compared to the case $g_X=\tilde g = 0$.
One sees that the enhancement in the Higgs mass when including loop effects and/or gauge-kinetic mixing is always smaller than the push at tree-level when neglecting these effects. Assuming a fixed ratio $\tilde g / g_X$, gauge kinetic mixing becomes more important for larger $g_X$. In contrast the (relative) impact of the loop corrections has only a mild dependence on the considered value of $g_X$. This can be seen in \cref{fig:Rel_T_L} where the shifts $\delta m^{(T)}$ and $\delta m^{(1L)}$ defined by
\begin{eqnarray}
\label{eq:d1}
\delta m_h^{(T)} &= & m_h^{(T)}(g_X) - m_h^{(T)}(g_X=0) \\
\label{eq:d2}
\delta m_h^{(1L)} &= &  m_h^{(1L)}(g_X) - m_h^{(1L)}(g_X=0) -\delta m_h^{(T)} 
\end{eqnarray}
are shown as function of $g_X$ and for different combinations of $M_{Z'}$ and $m_{\bar \eta}$. 
One finds, that the loop effects always reduce the overall enhancement of the Higgs mass by 25--30\%. The radiative corrections are usually more important for small $g_X$, but also depend to some extent on the hierarchy between $M_{Z'}$ and $m_{\bar \eta}$: for larger $m_{\bar \eta}$, which corresponds to a large enhancement at tree-level, the relative importance of the loops decreases faster with increasing $g_X$.

\section{Summary}
\label{sec:summary}
We have revisited here the possibility to push the tree-level mass of the SM-like Higgs by non-decoupling $D$-terms. It has been shown that a pure study at tree-level can overestimate the positive effect on the Higgs mass significantly. Namely, new radiative corrections triggered by the extended gauge sector can reduce the Higgs mass by several GeV. In addition, it has been shown that gauge kinetic mixing, which is a natural effect if the new Abelian gauge group is not orthogonal to $U(1)_Y$, reduces  the  positive shift of the Higgs mass at tree-level to some extent depending on the assumed cut-off scale $\Lambda$.

\section*{Acknowledgements}
I thank Toby Opferkuch for proof-reading the manuscript.


\begin{thebibliography}{10}
\bibitem{Chatrchyan:2012ufa}
{\bf CMS} Collaboration, S.~Chatrchyan {\em et.~al.}, {\it {Observation of a
  new boson at a mass of 125 GeV with the CMS experiment at the LHC}},  {\em
  Phys.Lett.} {\bf B716} (2012) 30--61
  [\href{http://arXiv.org/abs/1207.7235}{{\tt 1207.7235}}].

\bibitem{Aad:2012tfa}
{\bf ATLAS} Collaboration, G.~Aad {\em et.~al.}, {\it {Observation of a new
  particle in the search for the Standard Model Higgs boson with the ATLAS
  detector at the LHC}},  {\em Phys.Lett.} {\bf B716} (2012) 1--29
  [\href{http://arXiv.org/abs/1207.7214}{{\tt 1207.7214}}].

\bibitem{Nilles:1983ge}
H.~P. Nilles, {\it {Supersymmetry, Supergravity and Particle Physics}},  {\em
  Phys. Rept.} {\bf 110} (1984) 1--162.

\bibitem{Martin:1997ns}
S.~P. Martin, {\it {A Supersymmetry primer}},
  \href{http://arXiv.org/abs/hep-ph/9709356}{{\tt hep-ph/9709356}}. [Adv. Ser.
  Direct. High Energy Phys.18,1(1998)].

\bibitem{Batell:2015fma}
B.~Batell, G.~F. Giudice and M.~McCullough, {\it {Natural Heavy
  Supersymmetry}},  {\em JHEP} {\bf 12} (2015) 162
  [\href{http://arXiv.org/abs/1509.00834}{{\tt 1509.00834}}].

\bibitem{Camargo-Molina:2013sta}
J.~Camargo-Molina, B.~O'Leary, W.~Porod and F.~Staub, {\it {Stability of the
  CMSSM against sfermion VEVs}},  {\em JHEP} {\bf 1312} (2013) 103
  [\href{http://arXiv.org/abs/1309.7212}{{\tt 1309.7212}}].

\bibitem{Camargo-Molina:2014pwa}
J.~Camargo-Molina, B.~Garbrecht, B.~O'Leary, W.~Porod and F.~Staub, {\it
  {Constraining the Natural MSSM through tunneling to color-breaking vacua at
  zero and non-zero temperature}},  {\em Phys.Lett.} {\bf B737} (2014) 156--161
  [\href{http://arXiv.org/abs/1405.7376}{{\tt 1405.7376}}].

\bibitem{Chattopadhyay:2014gfa}
U.~Chattopadhyay and A.~Dey, {\it {Exploring MSSM for Charge and Color Breaking
  and Other Constraints in the Context of Higgs@125 GeV}},  {\em JHEP} {\bf
  1411} (2014) 161 [\href{http://arXiv.org/abs/1409.0611}{{\tt 1409.0611}}].

\bibitem{BasteroGil:2000bw}
M.~Bastero-Gil, C.~Hugonie, S.~F. King, D.~P. Roy and S.~Vempati, {\it {Does
  LEP prefer the NMSSM?}},  {\em Phys. Lett.} {\bf B489} (2000) 359--366
  [\href{http://arXiv.org/abs/hep-ph/0006198}{{\tt hep-ph/0006198}}].

\bibitem{Dermisek:2005gg}
R.~Dermisek and J.~F. Gunion, {\it {Consistency of LEP event excesses with an h
  ---> aa decay scenario and low-fine-tuning NMSSM models}},  {\em Phys. Rev.}
  {\bf D73} (2006) 111701 [\href{http://arXiv.org/abs/hep-ph/0510322}{{\tt
  hep-ph/0510322}}].

\bibitem{Ellwanger:2009dp}
U.~Ellwanger, C.~Hugonie and A.~M. Teixeira, {\it {The Next-to-Minimal
  Supersymmetric Standard Model}},  {\em Phys.Rept.} {\bf 496} (2010) 1--77
  [\href{http://arXiv.org/abs/0910.1785}{{\tt 0910.1785}}].

\bibitem{Ellwanger:2006rm}
U.~Ellwanger and C.~Hugonie, {\it {The Upper bound on the lightest Higgs mass
  in the NMSSM revisited}},  {\em Mod.Phys.Lett.} {\bf A22} (2007) 1581--1590
  [\href{http://arXiv.org/abs/hep-ph/0612133}{{\tt hep-ph/0612133}}].

\bibitem{Delgado:2010uj}
A.~Delgado, C.~Kolda, J.~P. Olson and A.~de~la Puente, {\it {Solving the Little
  Hierarchy Problem with a Singlet and Explicit $\mu$ Terms}},  {\em Phys. Rev.
  Lett.} {\bf 105} (2010) 091802 [\href{http://arXiv.org/abs/1005.1282}{{\tt
  1005.1282}}].

\bibitem{Ross:2011xv}
G.~G. Ross and K.~Schmidt-Hoberg, {\it {The Fine-Tuning of the Generalised
  NMSSM}},  {\em Nucl.Phys.} {\bf B862} (2012) 710--719
  [\href{http://arXiv.org/abs/1108.1284}{{\tt 1108.1284}}].

\bibitem{Hall:2011aa}
L.~J. Hall, D.~Pinner and J.~T. Ruderman, {\it {A Natural SUSY Higgs Near 126
  GeV}},  {\em JHEP} {\bf 04} (2012) 131
  [\href{http://arXiv.org/abs/1112.2703}{{\tt 1112.2703}}].

\bibitem{Ma:2011ea}
E.~Ma, {\it {Exceeding the MSSM Higgs Mass Bound in a Special Class of U(1)
  Gauge Models}},  {\em Phys.Lett.} {\bf B705} (2011) 320--323
  [\href{http://arXiv.org/abs/1108.4029}{{\tt 1108.4029}}].

\bibitem{Zhang:2008jm}
Y.~Zhang, H.~An, X.-d. Ji and R.~N. Mohapatra, {\it {Light Higgs Mass Bound in
  SUSY Left-Right Models}},  {\em Phys.Rev.} {\bf D78} (2008) 011302
  [\href{http://arXiv.org/abs/0804.0268}{{\tt 0804.0268}}].

\bibitem{Hirsch:2011hg}
M.~Hirsch, M.~Malinsky, W.~Porod, L.~Reichert and F.~Staub, {\it {Hefty
  MSSM-like light Higgs in extended gauge models}},  {\em JHEP} {\bf 1202}
  (2012) 084 [\href{http://arXiv.org/abs/1110.3037}{{\tt 1110.3037}}].

\bibitem{Lu:2013cta}
X.~Lu, H.~Murayama, J.~T. Ruderman and K.~Tobioka, {\it {A Natural Higgs Mass
  in Supersymmetry from NonDecoupling Effects}},  {\em Phys.Rev.Lett.} {\bf
  112} (2014) 191803 [\href{http://arXiv.org/abs/1308.0792}{{\tt 1308.0792}}].

\bibitem{Kaminska:2014wia}
A.~Kaminska, G.~G. Ross, K.~Schmidt-Hoberg and F.~Staub, {\it {A precision
  study of the fine tuning in the DiracNMSSM}},  {\em JHEP} {\bf 1406} (2014)
  153 [\href{http://arXiv.org/abs/1401.1816}{{\tt 1401.1816}}].

\bibitem{Ding:2015wma}
R.~Ding, T.~Li, F.~Staub, C.~Tian and B.~Zhu, {\it {The Supersymmetric Standard
  Models with a Pseudo-Dirac Gluino from Hybrid $F-$ and $D-$Term Supersymmetry
  Breakings}},  \href{http://arXiv.org/abs/1502.03614}{{\tt 1502.03614}}.

\bibitem{Batra:2003nj}
P.~Batra, A.~Delgado, D.~E. Kaplan and T.~M.~P. Tait, {\it {The Higgs mass
  bound in gauge extensions of the minimal supersymmetric standard model}},
  {\em JHEP} {\bf 02} (2004) 043
  [\href{http://arXiv.org/abs/hep-ph/0309149}{{\tt hep-ph/0309149}}].

\bibitem{Maloney:2004rc}
A.~Maloney, A.~Pierce and J.~G. Wacker, {\it {D-terms, unification, and the
  Higgs mass}},  {\em JHEP} {\bf 06} (2006) 034
  [\href{http://arXiv.org/abs/hep-ph/0409127}{{\tt hep-ph/0409127}}].

\bibitem{Babu:2004xg}
K.~Babu, I.~Gogoladze and C.~Kolda, {\it {Perturbative unification and Higgs
  boson mass bounds}},  \href{http://arXiv.org/abs/hep-ph/0410085}{{\tt
  hep-ph/0410085}}.

\bibitem{Martinez:2004rh}
R.~Martinez, N.~Poveda and J.~A. Rodriguez, {\it {Upper bound of the lightest
  higgs boson in a supersymmetric SU(3)l x U(1)x gauge model}},  {\em Phys.
  Rev.} {\bf D69} (2004) 075013.

\bibitem{Bertuzzo:2014sma}
E.~Bertuzzo and C.~Frugiuele, {\it {Natural SM-like 126 GeV Higgs boson via
  nondecoupling D terms}},  {\em Phys. Rev.} {\bf D93} (2016), no.~3 035019
  [\href{http://arXiv.org/abs/1412.2765}{{\tt 1412.2765}}].

\bibitem{Capdevilla:2015qwa}
R.~M. Capdevilla, A.~Delgado and A.~Martin, {\it {Light Stops in a minimal
  U(1)x extension of the MSSM}},  {\em Phys. Rev.} {\bf D92} (2015), no.~11
  115020 [\href{http://arXiv.org/abs/1509.02472}{{\tt 1509.02472}}].

\bibitem{Holdom:1985ag}
B.~Holdom, {\it {Two U(1)'s and Epsilon Charge Shifts}},  {\em Phys. Lett.}
  {\bf B166} (1986) 196.

\bibitem{Fonseca:2013bua}
R.~M. Fonseca, M.~Malinský and F.~Staub, {\it {Renormalization group equations
  and matching in a general quantum field theory with kinetic mixing}},  {\em
  Phys. Lett.} {\bf B726} (2013) 882--886
  [\href{http://arXiv.org/abs/1308.1674}{{\tt 1308.1674}}].

\bibitem{O'Leary:2011yq}
B.~O'Leary, W.~Porod and F.~Staub, {\it {Mass spectrum of the minimal SUSY B-L
  model}},  {\em JHEP} {\bf 05} (2012) 042
  [\href{http://arXiv.org/abs/1112.4600}{{\tt 1112.4600}}].

\bibitem{Staub:2016dxq}
F.~Staub {\em et.~al.}, {\it {Precision tools and models to narrow in on the
  750 GeV diphoton resonance}},  \href{http://arXiv.org/abs/1602.05581}{{\tt
  1602.05581}}.

\bibitem{Basso:2010jm}
L.~Basso, S.~Moretti and G.~M. Pruna, {\it {A Renormalisation Group Equation
  Study of the Scalar Sector of the Minimal B-L Extension of the Standard
  Model}},  {\em Phys. Rev.} {\bf D82} (2010) 055018
  [\href{http://arXiv.org/abs/1004.3039}{{\tt 1004.3039}}].

\bibitem{Hirsch:2012kv}
M.~Hirsch, W.~Porod, L.~Reichert and F.~Staub, {\it {Phenomenology of the
  minimal supersymmetric $U(1)_{B-L}\times U(1)_R$ extension of the standard
  model}},  {\em Phys. Rev.} {\bf D86} (2012) 093018
  [\href{http://arXiv.org/abs/1206.3516}{{\tt 1206.3516}}].

\bibitem{Krauss:2013jva}
M.~E. Krauss, W.~Porod and F.~Staub, {\it {SO(10) inspired gauge-mediated
  supersymmetry breaking}},  {\em Phys.Rev.} {\bf D88} (2013), no.~1 015014
  [\href{http://arXiv.org/abs/1304.0769}{{\tt 1304.0769}}].

\bibitem{Athron:2015vxg}
P.~Athron, D.~Harries, R.~Nevzorov and A.~G. Williams, {\it {$E_6$ Inspired
  SUSY Benchmarks, Dark Matter Relic Density and a 125 GeV Higgs}},
  \href{http://arXiv.org/abs/1512.07040}{{\tt 1512.07040}}.

\bibitem{delAguila:1988jz}
F.~del Aguila, G.~D. Coughlan and M.~Quiros, {\it {Gauge Coupling
  Renormalization With Several U(1) Factors}},  {\em Nucl. Phys.} {\bf B307}
  (1988) 633. [Erratum: Nucl. Phys.B312,751(1989)].

\bibitem{Fonseca:2011vn}
R.~M. Fonseca, M.~Malinsky, W.~Porod and F.~Staub, {\it {Running soft
  parameters in SUSY models with multiple U(1) gauge factors}},  {\em
  Nucl.Phys.} {\bf B854} (2012) 28--53
  [\href{http://arXiv.org/abs/1107.2670}{{\tt 1107.2670}}].

\bibitem{Basso:2010pe}
L.~Basso, A.~Belyaev, S.~Moretti, G.~M. Pruna and C.~H.
  Shepherd-Themistocleous, {\it {$Z'$ discovery potential at the LHC in the
  minimal $B-L$ extension of the Standard Model}},  {\em Eur. Phys. J.} {\bf
  C71} (2011) 1613 [\href{http://arXiv.org/abs/1002.3586}{{\tt 1002.3586}}].

\bibitem{Krauss:2012ku}
M.~E. Krauss, B.~O'Leary, W.~Porod and F.~Staub, {\it {Implications of gauge
  kinetic mixing on Z' and slepton production at the LHC}},  {\em Phys. Rev.}
  {\bf D86} (2012) 055017 [\href{http://arXiv.org/abs/1206.3513}{{\tt
  1206.3513}}].

\bibitem{Basso:2012gz}
L.~Basso, B.~O'Leary, W.~Porod and F.~Staub, {\it {Dark matter scenarios in the
  minimal SUSY B-L model}},  {\em JHEP} {\bf 09} (2012) 054
  [\href{http://arXiv.org/abs/1207.0507}{{\tt 1207.0507}}].

\bibitem{Porod:2011nf}
W.~Porod and F.~Staub, {\it {SPheno 3.1: Extensions including flavour,
  CP-phases and models beyond the MSSM}},  {\em Comput. Phys. Commun.} {\bf
  183} (2012) 2458--2469 [\href{http://arXiv.org/abs/1104.1573}{{\tt
  1104.1573}}].

\bibitem{Porod:2003um}
W.~Porod, {\it {SPheno, a program for calculating supersymmetric spectra, SUSY
  particle decays and SUSY particle production at e+ e- colliders}},  {\em
  Comput. Phys. Commun.} {\bf 153} (2003) 275--315
  [\href{http://arXiv.org/abs/hep-ph/0301101}{{\tt hep-ph/0301101}}].

\bibitem{Staub:2008uz}
F.~Staub, {\it {SARAH}},  \href{http://arXiv.org/abs/0806.0538}{{\tt
  0806.0538}}.

\bibitem{Staub:2009bi}
F.~Staub, {\it {From Superpotential to Model Files for FeynArts and
  CalcHep/CompHep}},  {\em Comput.Phys.Commun.} {\bf 181} (2010) 1077--1086
  [\href{http://arXiv.org/abs/0909.2863}{{\tt 0909.2863}}].

\bibitem{Staub:2010jh}
F.~Staub, {\it {Automatic Calculation of supersymmetric Renormalization Group
  Equations and Self Energies}},  {\em Comput.Phys.Commun.} {\bf 182} (2011)
  808--833 [\href{http://arXiv.org/abs/1002.0840}{{\tt 1002.0840}}].

\bibitem{Staub:2012pb}
F.~Staub, {\it {SARAH 3.2: Dirac Gauginos, UFO output, and more}},  {\em
  Comput.Phys.Commun.} {\bf 184} (2013) pp. 1792--1809
  [\href{http://arXiv.org/abs/1207.0906}{{\tt 1207.0906}}].

\bibitem{Staub:2013tta}
F.~Staub, {\it {SARAH 4: A tool for (not only SUSY) model builders}},  {\em
  Comput.Phys.Commun.} {\bf 185} (2014) 1773--1790
  [\href{http://arXiv.org/abs/1309.7223}{{\tt 1309.7223}}].

\bibitem{Staub:2015kfa}
F.~Staub, {\it {Exploring new models in all detail with SARAH}},
  \href{http://arXiv.org/abs/1503.04200}{{\tt 1503.04200}}.

\bibitem{Goodsell:2014bna}
M.~D. Goodsell, K.~Nickel and F.~Staub, {\it {Two-Loop Higgs mass calculations
  in supersymmetric models beyond the MSSM with SARAH and SPheno}},  {\em
  Eur.Phys.J.} {\bf C75} (2015), no.~1 32
  [\href{http://arXiv.org/abs/1411.0675}{{\tt 1411.0675}}].

\bibitem{Goodsell:2015ira}
M.~Goodsell, K.~Nickel and F.~Staub, {\it {Two-loop Higgs mass calculation from
  a diagrammatic approach}},  \href{http://arXiv.org/abs/1503.03098}{{\tt
  1503.03098}}.

\end{thebibliography}
\end{document}